# A Novel and Efficient Data Point Neighborhood Construction Algorithm based on Apollonius Circle

Shahin Pourbahrami[1], Leyli Mohammad Khanli[*,1], Sohrab Azimpour[2]

Computer Engineering Department, Faculty of Electrical and Computer Engineering, University of Tabriz, Tabriz, Iran[1]
Mathematic Department, Faculty of Mathematic, University of Farhangian, Tehran, Iran[2]

**Abstract**

Neighborhood construction models are important in finding connection among the data points, which helps demonstrate interrelations among the information. Hence, employing a new approach to find neighborhood among the data points is a challenging issue. The methods, suggested so far, are not useful for simultaneous analysis of distances and precise examination of the geometric position of the data as well as their geometric relationships. Moreover, most of the suggested algorithms depend on regulating parameters including number of neighborhoods and limitations in fixed regions. The purpose of the proposed algorithm is to detect and offer an applied geometric pattern among the data through data mining. Precise geometric patterns are examined according to the relationships among the data in neighborhood space. These patterns can reveal the behavioural discipline and similarity across the data. It is assumed that there is no prior information about the data sets at hand. The aim of the present research study is to locate the precise neighborhood using Apollonius circle, which can help us identify the neighborhood state of data points. High efficiency of Apollonius structure in assessing local similarities among the observations has opened a new field of the science of geometry in data mining. In order to assess the proposed algorithm, its precision is compared with the state-of-the-art and well-known (k-Nearest Neighbor and epsilon-neighborhood) algorithms.

**Keywords:** Apollonius circle, geometric patterns, neighborhood construction

## 1. Introduction

Neighborhood includes a group of data points, which are locally similar to each other, being defined according to their local inter-relationships in a database (İnkaya et al., 2015). One of the main features of data mining is analysis of the data and their categorization in similar groups by a precise examination of the neighborhood features of the data. Neighborhood is a criterion to make precise neighborhood models by changing the data tendencies towards similar categories. Similarity, here it refers to the intrinsic dependency inside the data and automatic connection of similar sets of data. In other words, neighborhood for each set of data is defined in terms of similarity and connection among the data in such a way that the obtained precision can satisfy the current challenges for data mining in different areas. In fact, the purpose is to discover the hidden information and the interconnections in the current data and to predict the unclear or unobserved cases.

The *k*-Nearest Neighbor (*k*-NN) is one of the most commonly used algorithms in defining neighborhood's data (graph-based) and clustering them, according to the Euclidean distance. First, it decides to select the similar data sets. Then, the clusters are determined (Qin et al., 2018). Although this algorithm is simple and effective, determining neighborhood in this algorithm depends only on the distance and geometric location of the points, and statistical rules are not considered. Meanwhile, parameter *k* has an important role in defining neighborhood (Stork et al., 2001, Maillo et al., 2017, Pan et al., 2015, Güney & Atasoy, 2012, Mohammadi et al., 2015, García-Pedrajas et al., 2017).

The simple epsilon (ε) method determines neighborhood according to a small radius and the parameter of epsilon distance (Pedrycz, 2010). In this area of neighborhood, the data are clustered based on parameter of epsilon. If an inappropriate parameter is selected, the efficiency epsilon of the approach will be decreased. Therefore, this method lacks a strong and precise neighborhood structure. In this method, if the value of the selected epsilon is small, the concerned point in the assigned radius may have no neighbors, and as a result, it may be identified as outlier data while not being actually the

Corresponding Author. [*]
*Email Addresses:* sh.pourbahrami@tabrizu.ac.ir (Shahin Pourbahrami), l-khanli@tabrizu.ac.ir (Leyli Mohammad Khanli), azimpour@cfu.ac.ir (Sohrab Azimpour).

case. If neighborhood radius is selected to be high, different points in different groups may integrate into one group.

Therefore, a novel locally neighborhood construction algorithm is proposed to address such limitations of the neighborhood approaches. The main idea of our algorithm is to determinate the geometric relations between data points in data sets by Apollonius circles (Partensky, 2008, Hoshen, 1996). In order to show the performance of our algorithm, it can be applied to clustering problems. It is assumed that there is no prior information about the data set at hand. The advantages of our algorithm are local outliers detection and no fix neighborhood region to optimize the efficiency of finding neighborhood data points. The algorithm proposed for forming Apollonius circle by using geometric structure, without any need for examining the distance of each individual point from the appointed center, forms similar groups. This algorithm identifies the outliers using geometric structure. In neighborhood formation algorithms, KNN and Neighborhood Construction (NC) should be examined for each individual point in neighborhood data set so that similar groups would be formed (İnkaya et al., 2015). Therefore, the use of Apollonius circle for neighborhood formation, offers a suitable structure for geometric separation of the defined center from the neighboring points to the next nearest center so that the most similar points would be located within the Apollonius circles defined for them. Neighborhood structures constructed by our algorithm work well where (1) different clusters have different densities, (2) density varies within a cluster, and (3) extraction neighborhood structure in assessing the local similarities among the data points. The proposed algorithm was utilized since interesting orbital connections could be extracted for locating neighbourhood among the points under investigation (attraction of local nearby points); it could also reduce the complexity of the previous algorithms. Moreover, the proposed algorithm is able to explore the outlier data in different noisy data sets. The algorithm proposed not need be examined for each individual point in neighborhood data set. The main limitation of our proposed algorithm is low accuracy in high-dimension data sets.

The development of intelligent and expert systems is and has been one of the priorities of researchers to provide aspects the diagnosis of diseases (Bautista, et al., 2018).The high efficiency of Apollonius structure in assessing the local similarities among the observations has opened a new field of the science of geometry in data mining. For example, Apollonius structure will be using for medical decision support system to predict disease in future. Apollonius structure allows for the representation of imprecise knowledge via the introduction of geometry measures, providing a powerful method of similarity description among instances. In Apollonius structure methods, geometry set theories are introduced into neighborhood, which assign membership degrees to different classes instead of the distances to their nearest neighbors.

The rest of the paper is organized as follows. Section 2 provides an overview of the key related works on neighborhood construction. In Section 3, steps of our algorithm are discussed which are used for neighborhood construction in this paper. In Section 4, the assessment criteria are discussed. In section 5, our proposed algorithm is applied to several artificial and real world data sets and the results and discussion are presented, afterward. Finally, Section 6 explains the conclusions of the paper.

## 2. Review of Related Literature

Several other methods have been offered to improve classification and clustering precision, using graphs and relationship neighborhoods. Graphs such as Gabriel's graph are geometric methods to examine the direct and indirect relationships among the points (Varma et al., 2016, İnkaya, 2015a, İnkaya, 2015b, Güngör & Özmen, 2017). As shown in Fig.1, to examine the relationship between two points, Gabriel's graph determines the distance between the two points of $x_i, x_q$ as the circle diameter. If there are no other points inside Gabriel's neighborhood circle, it shows that the two mentioned points have a direct relationship and direct neighborhood. Otherwise, it extracts the number of the points inside Gabriel's circle as the density between the two points.

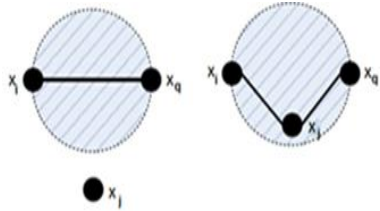

**Fig. 1.** Gabriel graph, İnkaya (2015b).

NC algorithm is based on the geometric structure and density between the points (İnkaya et al., 2015). In the process of dealing with NC algorithm, after a single point, all the other points are arranged from the nearest to the farthest ones. Then, using Gabriel's graph, the direct and indirect connections for all points in the database are extracted, and the densities of direct and indirect points are saved in a set. If the density inside the set or the densities of the points is decreased (decrease point is called break point), a new neighborhood group is constructed. In the new neighborhood groups between the two sets, precise examination is carried out on the neighborhood, which is the intersection between the break points and the neighborhood group of a special point. If there is an intersection between the neighbors of these points, they will also be considered as neighbors of each other. Otherwise, each of them will form a separate neighborhood group. In the last phase of the algorithm, the mutual relationship of the set of nearest neighbors is examined to determine the outlier data. The outlier data are also extracted and identified. Since this algorithm is suitable for the data sets in small databases, it may have a high complexity in the big data sets.

In Fig.2 (a), Relative Neighborhood Graph (RNG) is displayed (Yang et al., 2016). In RNG, $x_i, x_q$ are determined as the centers of two circles, and the distance between $x_i, x_q$ are the radius of the circles. If there are no other points inside, these two points are considered as neighbor points. However, the problem of this method and Gabriel's graph is that they have fixed number neighbors; which decreases in the optimization of finding neighborhood points. This is a disadvantage of RNG and GG. The "disadvantage" means that the number of neighbors are fixed for a given query sample, therefore, it is impossible to make further the optimization of clustering performance.

In geometric graphs theory, $\beta$-skeleton is the skeleton of an undirected graph, which is defined on a set of Euclidean geometric points (Yang et al., 2016). In $\beta$-skeleton method, the value of $\beta$ should be identified in order to determine neighborhood area. In this method, based on $\beta$ parameter for two points of $x_i, x_q$ using (1), angle $\theta$ is determined. With the obtained angle, neighborhood for the angles of more than 90 degrees is the intersection region between two points of the two circles of the subtended arc. For the angles of less than 90 degrees, the union of two circles of subtended arc is considered as the neighborhood. If the angle is 90 degrees, the method performs like Gabriel's method. Fig.2 (b) shows performance of this method with different $\beta$ values. In $\beta$-skeleton method, if the value of $\beta$ is more than *1*, the graph neighboring area is considered as the nearest neighborhood. In $\beta$-skeleton method, $\beta$ parameter needs to be regulated (Cardinal et al., 2009, Veltkamp, 1992, Yang et al., 2016). Gabriel's graph methods and RNG do not need to regulate (tune) their parameters. $\beta$-skeleton has a parameter of $k$, based on which, the size and shape of the graph can be changed. This makes the further optimization possible; however, there exists the problem of parameter selection.

$$\theta = \begin{cases} \sin^{-1}\dfrac{1}{\beta} & if \quad 1 \leq \beta \\ \pi - \sin^{-1}\beta & if \quad \beta \leq 1 \end{cases} \quad (1)$$

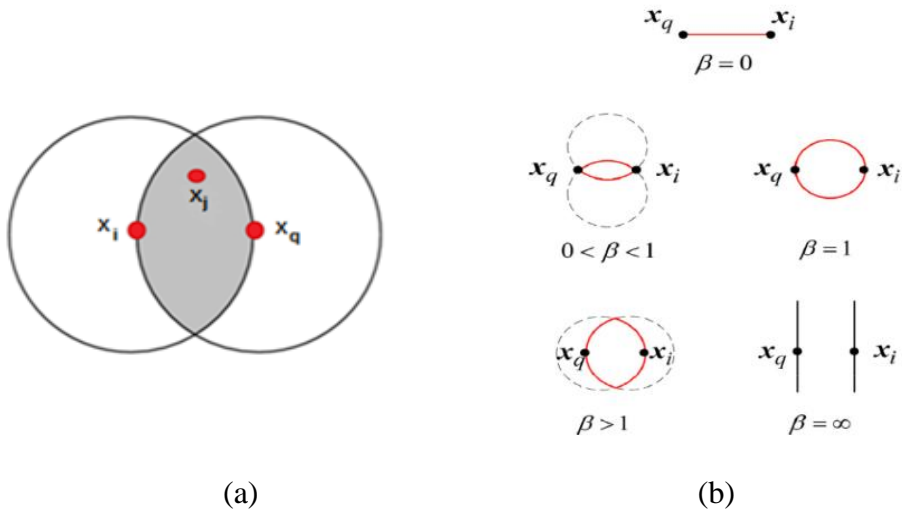

**Fig. 2** (a), RNG and (b), β-skeleton, Yang et al., (2016).

Another method called Angle-based Neighborhood Graph (ANG) finds neighborhood of data points according to the angle between points (Yang et al., 2016). This method uses geometric relationships and by defining the parameter angle, extracts the angle between direct and indirect geometric points. The angle $\theta$ between $x_i, x_q$ is determined on the circle of subtended arc and if the angle between $x_i, x_j, x_q$ is smaller than the angle between $x_i, x_q$, these the two points have a direct relationship with each other. The disadvantages of this method include regulation of the angle parameter and its complicated calculations; the advantage is the size of neighborhood which can be adjusted according to the angle, and this size is efficient in determining the optimal precision in the data point neighborhood.

Central Diagram Method uses visualization to show direct and indirect relationships among the nodes; which distributes the nodes on the circles which have central nodes, and defines different communication layers to determine neighborhoods. The centers of the circles include two basic nodes which can be examined in the database. The relationships of the other visualized nodes are determined according to these centers. The most inner layers show direct or near connections, but most outer layers indicate farther connections (indirect connections or no connection). In this way, the direct and indirect nodes are extracted (Park, & Basole, 2016, Crnovrsanin et al., 2014, Vehlow et al., 2015).

Optimum Path Forest ($OPF_{kNN}$) is a method to find data point neighborhood using graphs (Papa et al., 2017). In its first phase, this method uses the distance between the nodes and forms complete graphs. In its second phase, it obtains spanning tree of the data and identifies the dense nodes. Then, it estimates the cost for each node (the cost of the distance between two nodes, zero cost for the dense nodes themselves, and infinite cost for the normal nodes). In training phase, it finds the data point neighborhood by minimizing and maximizing the costs, and in the testing phase, according to minimal or maximal rate of group costs, conductive of the test node is identified by the groups.

*K*-Associated Optimal Graph determines the number of neighborhoods obtained from *k*-NN algorithm of neighborhood. Starting with *1*, it increases the number of *ks* in each phase for groups obtained from the nodes and calculates the purity rate (It is calculated according to the set of input and output nodes and their value in each group). In each phase of increasing value *k*, the groups may integrate with each other (purity is recalculated), and if purity increases again, a suitable combination is obtained. Otherwise, the two groups do not integrate and stay in their previous state. The problem with this algorithm is its high complexity in big databases (Mohammadi et al., 2015).

Self-Organizing Map (SOM) is a method in which competitive learning method is used to train and has been developed based on specific features of human brain (Vesanto & Alhoniemi, 2000). SOM projects a high-dimensional data set on the prototype vectors of a low-dimensional grid structure. Each grid is related to a neuron. SOM network is trained to assign similar data points to the same neuron. Two neighboring neurons on the grid imply close data points in the original data set. Therefore, SOM helps to visualize and probe neighborhood relations in the data set (Hajjar & Hamdan, 2013, Awad, 2010).

The idea behind Density Peaks Clustering (DPC) is that cluster centers are denoted by a higher density than their neighbors and by a relatively large distance from data points with higher densities (Rodriguez & Laio, 2014, Li & Tang, 2018). The two weaknesses of DPC are not sensitive to local geometric and do not perform well when data points have relatively high dimensions. In DPC-KNN after recognition cluster centers, each data point is assigned to its nearest center (Rodriguez & Laio, 2014). The DPC-PCA is an algorithm based on principal component analysis that reduces the data and keeps 99% of principal components to eigenvectors (Du et al., 2016).

*k*NN algorithm is regarded as a simple and most commonly used method due to its high comprehensibility and not requiring hypothesis formation on the data. Despite its advantages, this algorithm suffers from two problems to be discussed below: The first problem with this algorithm concerns the determination of *k* value by the user which is highly important for extracting connections and local features among the points. The second problem lies in the fact that in determining neighborhood, it is also highly important to decide on the criterion of similarity among the points (Qin et al., 2018). The problem with NC algorithm is its high complexity since this algorithm is suitable for the data sets in small databases. NC is especially useful for their accuracy in locating neighborhood points with the highest rate of similarity based on efficient Gabriel structures. It can increase accuracy for low dimension of data points in grouping similar data. NC should be examined for each individual point in neighborhood data set so that similar groups would be formed (İnkaya et al., 2015). DPC-KNN does a poor job of finding the clusters of high-dimensional data. It may generate wrong number of clusters of real-world data sets. This is because many of the dimensions in high dimensional data are often irrelevant. These irrelevant dimensions can confuse DPC by hiding clusters in noisy data. Another reason is its overwhelming dependency on the distances between points.

## 3. Neighborhood Construction by Apollonius Region (NCAR) Algorithm

Because of the significance of precision in finding neighborhood data points, the proposed algorithm focuses on offering an accurate data point neighborhood construction which removes the shortcomings of the previous models such as fixed neighborhood region and high complexity. In addition, the algorithm can identify similar groups and prevent finding imprecise data point neighborhoods (outlier data can be detected).

For the mentioned problems we have proposed one algorithm to determine the neighborhoods which do not need fixed neighborhood region. To overcome the problems of the previous algorithms, Apollonius structure is used as a very well-known problem in geometry (Partensky, 2008, Hoshen, 1996). The newly proposed algorithm of constructing data point neighborhood is called Neighborhood Construction by Apollonius Region (NCAR) and has wide applications in geometry-based data point and data mining. In these set of points, there is no prior information about the data set.

The approach proposed in this algorithm has better prospect to improve precision of neighborhood construction in comparison with Gabriel's graph and nearest neighborhood graph. Instead of determining neighborhood for just one point, it decides for two points simultaneously. This algorithm is useful in finding community in social networks, edge bundling in information visualization, and clustering the data in wireless sensor networks (Pourbahrami, S. 2017); it is also useful in determining the nearest and the most similar neighbors of a test sample. Besides, it comprehensively uses the geometric information and distance between the points.

### 3.1 Notation and definition

We use the notation given below throughout our algorithm in Table 1.

**Table 1**
The notation of NCAR algorithm.

| | |
|---|---|
| M | set of data points |
| i,j,l,n,m,t | indices for data points |
| T | set of target points (high density) |
| $(T_t, T_{t+1})$ | list pair target point in ascending order |
| $FP$ | farthest data point |
| $Fd$ | Farthest distance |
| $C_{AB}$ | Apollonius circle for points A and B |
| $C_B$ | Apollonius circle side point B |
| $C_A$ | Apollonius circle side point A |
| $C_{inf}$ | Apollonius circle at infinet |
| $R_{C_{AB}}$ | radius Apollonius circle |
| $C_{C_{AB}}$ | center Apollonius circle |
| $\mathbb{P}$ | Euclidean plane |
| d(A, M) | Euclidean distance between points A and M |
| k | distance ratio of (A, M) over (M, B) |

**Definition of $C_{AB}$:** Apollonius circle is the geometric position of the points in Euclidean plane where the distance of the points from the assumed fixed points of *A* and *B* is a fixed value of *k* (*k*≠0) Eq. (2) (Partensky, 2008, Hoshen, 1996).

$$k = \frac{d(A,M)}{d(M,B)} \qquad (2)$$

In three dimensional spaces, the Apollonius circle turns into Apollonius sphere which is defined in the following way:

**Definition of Apollonius sphere:** Suppose *k* is a constant and positive number. Moreover, *A* and *B* are two points in the space, the geometric position of the points in the space that apply in equation 2 would be a sphere called Apollonius sphere.

The Apollonius circles of points *A* and *B* are shown with $C_{AB}$ in Eq. (3) (Shown in Fig. 3).

$$C_{AB} = \begin{cases} C_A & \text{if } k < 1 \\ C_B & \text{if } k > 1 \\ C_{\inf} & \text{if } k = 1 \end{cases} \qquad (3)$$

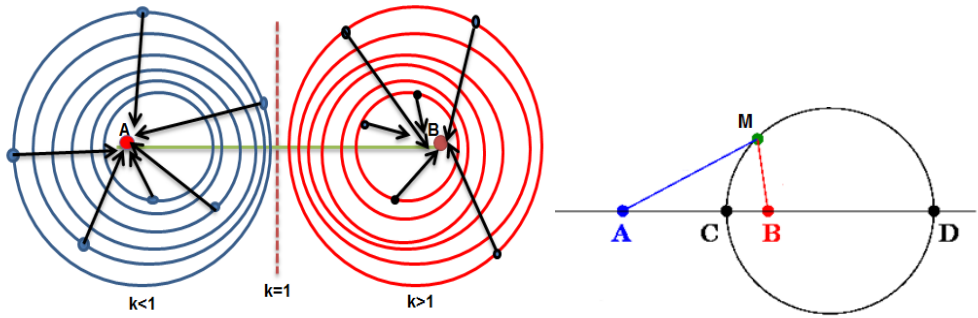

**Fig. 3.** Apollonius circles $C_{AB}$ and $C_B$

**Proof by geometric method**: Suppose the two fixed points of *A* and *B* on plane $\mathbb{P}$, draw straight line of AB, and take point C and D on this line in such a way that it divides line *AB* by *k* or $\frac{CA}{CB} = \frac{DA}{DB} = k$. The circle with a diameter of CD is the respective geometric position. In other words, geometric position is the point whose distance with two fixed points of A and B is *k*. Actually, the distance of any point like M which is located on this circle is *k* if M is connected to A, B, C, and D (for ABCD is a conventional division), then M-ABCD is a conventional coordinate. Moreover if the two non-subsequent radiuses of this conventional coordinates or MC and MD are perpendicular to each other ( $\angle CMD$ is an inscribed angle opposite the diameter and is a 90˚angle), then these two radiuses are bisectors of the internal and external angles between the two other radiuses. In other words, MC is the bisector of the internal angle of $\angle AMB$, and MD is the bisector of the external angle of $\angle AMB$. On the other hand, the bisector of each angle divides the opposite side it into two smaller sides.

**Definition of C (Apollonius Circle Side B and A)**: If the value of *k* becomes more than *1*, Apollonius circle is drawn on the side of point (data) *B* shown in Fig. (4). In this case, *B* is located inside the Apollonius circle, so, the neighborhood of *B* will be specified. If $0 < k < 1$, then Apollonius circle is drawn on the side of point *A* shown in Fig. 5.

Note: The higher the value of *k* (the value increase to more than *1*) the nearer Apollonius circle gets to point *B*. Moreover, the less the value of *k* (the nearer it is to *1*), the larger the Apollonius circle becomes. Apollonius Circle $C_B$ is defined by Eq. (4), where [*x, y*] is the geometrical location of the points *A, B* on Apollonius circle in *x*-axis and *y*-axis. In equation $C_B$, if $k \to \infty$ $C_B$ gets tighter, and if

$k \to 1$, $C_B$ gets larger. Apollonius Circle $C_A$ is defined by Eq. (5). In equation $C_A$, if $k \to 0$ $C_A$ is tighter, and if $k \to 1$, $C_A$ is larger.

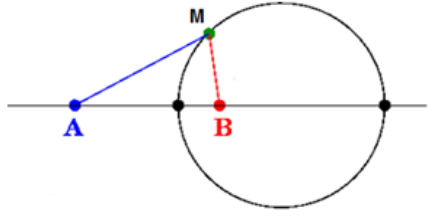

**Fig. 4.** Apollonius circle $C_B$

$$C_B = \begin{cases} \dfrac{k\sqrt{(x_B-x_A)^2+(y_B-y_A)^2}}{|1-k^2|} \to 0 & k \to \infty \\ \dfrac{k\sqrt{(x_B-x_A)^2+(y_B-y_A)^2}}{|1-k^2|} \to \infty & k \to 1^+ \end{cases} \quad (4)$$

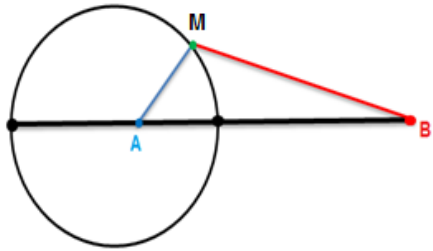

**Fig. 5.** Apollonius circle $C_A$

$$C_A = \begin{cases} \dfrac{k\sqrt{(x_B-x_A)^2+(y_B-y_A)^2}}{|1-k^2|} \to 0 & k \to 0 \\ \dfrac{k\sqrt{(x_B-x_A)^2+(y_B-y_A)^2}}{|1-k^2|} \to \infty & k \to 1^- \end{cases} \quad (5)$$

**Definition of $C_{inf}$ (Apollonius Circle at Infinite)**: if $k=1$, M is located on perpendicular bisector of the line segment A and B, and Apollonius circle turns into a straight line which the center is located in infinite (shown in Fig. 6).

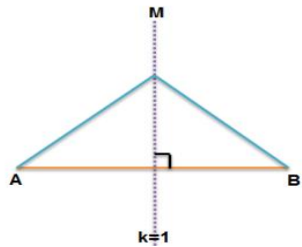

**Fig. 6.** Apollonius circle $C_{inf}$

## 3.2 Steps of the NCAR algorithm

The NCAR algorithm determines the neighbors for each data point. The NCAR algorithm consists of three consecutive steps. In the first step, the proposed algorithm extracts high density data points. In the second step, neighborhood groups by drawing Apollonius circles are determined. In the third step, the neighborhoods among Apollonius circles are verified.

**Step 1: Extraction of the points with high density**

Rodrigues and Laio proposed an algorithm for extraction of high density data points which is applied using Eq. (6) to (9). In this method, the high density data points are extracted and are saved in a list (Wang et al., 2016, Rodriguez & Laio, 2014, Du et al., 2016). In this paper, we will refer them as target points. Let's assume that $M = \{M_{1i}, M_{2i}, ..., M_{mi}\}$ is a data point with $m$ features, and $n$ is the number of data points, where $N_{M_i}$ shows $k$ nearest neighbors of data point $M_i$. Where $d(M_i, M_j)$ shows the Euclidean distance between data point $M_i$ and $M_j$. The parameter $r$ which was used in density based clustering algorithms refers to a portion of data points, which number of the neighbors (i.e. $r = p \times n$ and $r$ is the number of neighbors). The local density $\rho_i$ for data point $M_i$ is calculated as follows:

$$\rho_i = \exp\left(-\left(\frac{1}{r} \sum_{M_j \in N(M_i)} d(M_i, M_j)^2\right)\right) \tag{6}$$

$$d(M_i, M_j) = \|M_i - M_j\| \tag{7}$$

Let's $\delta_i$ denote the minimum distance between point $M_i$ and any other higher density points. This measure is denoted by $\delta_i$ and is calculated as follows:

$$\delta_i = \begin{cases} \min_{j:\rho_i < \rho_j} \{d(M_i, M_j)\}, & \text{if } \exists j \text{ s.t. } \rho_i < \rho_j \\ \max_j \{d(M_i, M_j)\}, & \text{otherwise} \end{cases} \tag{8}$$

In Eq. (9), if the value of the two obtained parameters $\delta_i$ and $\rho_i$ are rather high, the groups are formed according to the determined target points. Score function is calculated to rank the data points:

$$score(M_i) = \delta_i \cdot \rho_i \tag{9}$$

**Step 2: Construction of neighborhood groups with Apollonius circle**

In this section, there are three phases. In the first phase, the closest target points are considered as fixed points (*A, B*) in Apollonius circle. In the second phase, the farthest point from each target point is determined. In the third phase, Apollonius circles are drawn.

**The first phase**: After finding target points, they are sorted according to their least distance from each other (the number of target points and Apollonius circle is equal). In the first phase of the second step these points are considered as fixed hypothetical points in Apollonius circle. The target points are shown with $T = \{T_1, T_2, ..., T_m\}$ that Sorted $(T_t, T_{t+1})$ and $t \in \{1, 2, ..., m-1\}$. The data points are shown with $M = \{M_i | i \in \{1, 2, ..., n-m\}, M_i \notin T, M \cap T = \emptyset\}$.

Note: It is important to sort pair circle target points according to the distances so that they would not overlap.

**The second phase:** In Eq. (10) and Eq. (11), we find the farthest distance ($Fd$) from the target points, and all points that are nearer to the target point than the farthest point (FP), are saved in a list. The farthest point to a target point (*B*) is determined in such a way that its distance to *B* is less than the distances between the furthest point and all other targets. The distance of the farthest point with a target point should be less than distance between two target points (to detect outliers).

$$Fd_{T_t} = \max \begin{Bmatrix} d(T_t, M_i) | M_i \in M \text{ and } d(T_t, M_i) < d(T_t, T_{t+1}) \text{ and} \\ d(T_t, M_i) < \min_{l=1, l \in T}^{m} d(T_l, M_i) \text{ s.t. } t \neq l \end{Bmatrix} \quad (10)$$

$$FP_{T_t} = \{M_i \mid d(T_t, M_i) = Fd_t\} \quad (11)$$

**The third phase:** In this phase, the purpose is grouping data points with Apollonius circle according to the farthest point and target point. In other words, target points are connected to be point *A* and *B* in the definition of Apollonius circle, and to find point *M* on Apollonius circle, the farthest point from the first target point (which is even farther than the other target points) is selected. Then, according to the distance between the points, total *k* value is obtained, and Apollonius circle is drawn. In Eq. (12) and Eq. (13) according to target points and value *k*, radius and center needed for drawing Apollonius circle is obtained.

$$R_{C_{AB}} = \frac{k\sqrt{(x_B - x_A)^2 + (y_B - y_A)^2}}{|1 - k^2|} \quad (12)$$

$$C_{C_{AB}} = \left( \frac{(x_B - k^2 x_A)}{|1 - k^2|}, \frac{(y_B - k^2 y_A)}{|1 - k^2|} \right) \quad (13)$$

This continues for the other points in the target points list until groups related to all target point are constructed. It should be noted that this phase of algorithm is used to reduce the complexity and determine the primary groups by using Apollonius circle (it divides *AB* harmonically and orbitals). In the NCAR algorithm, time complexity is lower than other geometry-based neighborhood construction algorithms, because there is not neighborhood relation computation for each point, and the neighborhood of a point need not be compared to those of all the points with Apollonius's geometry.

**Step 3: Verifying to neighborhood between Apollonius circles**

In this section, step 3 has two phases. The first phase is to find data points out of radius Apollonius circles or data points on $C_{inf}$ circles. In the second phase, data points are extracted inside overlap of Apollonius circles. Then, we reassign them to their most similar neighbors.

**The first phase:** The nearest extracted points in the second step for each target point is considered according to the radius of the Apollonius circle. If the distance of these points from the center of Apollonius circle is greater than the radius of the Apollonius circle or data points on circles $C_{inf}$ (k = 1), they take the label of their nearest target points which contain the greatest number of neighborhoods as well. This is done by averaging distance of the mentioned points and the points inside the circles and reassigning it to the center of Apollonius circles with shortest distance or by using similarity measure for reassigning points to their most similar sets.

**The second phase:** The points falling inside the overlap area of the Apollonius circles are also extracted and are reassigned to any of the neighbors in the two or more circles to which they are most similar, or else they are reassigned to the nearest center of Apollonius circles. The pseudo-code steps of the NCAR algorithm are shown in Fig.7.

**Algorithm:** NCAR Pseudo-code

**Inputs**: The Samples M ∈ $R_{N \times M}$
The Parameter P
**Outputs**: The label vector of group index: y ∈ $R_{N \times 1}$
**Method:**
**Step 1-** Identifying target points
1: Calculate distance matrix using Eq. (1)
2: Calculate $\rho$ for each point using Eq. (2)
3: Calculate $\delta$ for each point using Eq. (3)
4: Calculate the score for each point using Eq. (4)
5: Plot decision graph and select target points
**Step 2**: Forming groups with Apollonius circle
1: Assign points of *A*, *B* circle Apollonius to set target points
2: Calculate the distance of point *M* (further of target points) for each Apollonius circle or target point
3: Grouping based on Apollonius circle and assigning sets of points to target points
**Step 3**: Reassign points out of the radius of Apollonius circle or on circle $C_{inf}$
1: Reassign each remaining point out of radius circle Apollonius to the new group based on the most similar neighbors and the least distance to the center of Apollonius circles.
2: Reassign each remaining points in overlap between two Apollonius circles to the new group based on the most similar neighbors and the least distance to the center of Apollonius circles
**Return y**

**Fig. 7.** Pseudo-code the NCAR algorithm

In the example Fig.8 (a)-(d) given below, out of the specified points, point 1, 5 and 8 have the target points which have been obtained using first step formulas, ($T = \{1, 5, 8\}$ Sorted (1,5)) and Apollonius circles were also drawn by the method in the second step (point 4 is found by the algorithm as the farthest point 1 which is also away from point 8 and 5, for point 1, $FP_1 = 4$, G1 = {2, 3, 4}, and points 6 and 7 are next farthest points of 5 and 8). Then Apollonius circle is drawn for forming Groups = {G1, G2, G3}. In this phase, the outlier data are distant from the Apollonius circle and farther away from the data points inside Apollonius circles (point 10). If for one target point, point FP is obtained as the most distant point from the other target points, but its distance from the target points is more than distance between two target points, that point has potential of having outlier data and is introduced as the outlier data in the algorithm.

$$Fd_1 = \max \begin{Bmatrix} d(1, M_i) | M_i \in \{2,3,4,6,7,9,10\} \text{ and } d(1, M_i) \langle d(5, M_i) \text{ and} \\ d(1, M_i) \langle d(8, M_i) \text{ and } d(1, M_i) \langle d(1,5) \end{Bmatrix} = d(1, 4) \rightarrow FP_1 = 4$$

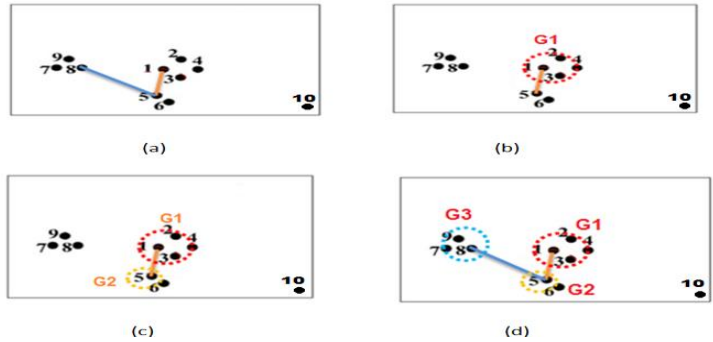

**Fig. 8.** Example data set for the NCAR algorithm, (a) points 1, 5 and 8 are target points, (b) point 4 is the farthest point, formation G1, (c) point 6 is the other farthest point, formation G2, (d) point 7 is other the farthest point, formation G3.

### 3.3. Theoretical background of the NCAR algorithm

In this section, we present the theory related of NCAR algorithm, as well as the basis of NCAR-based neighborhood construction. In using Apollonius circle, the value of *k* has a determining role in examining the accuracy of grouping points (observations) and their attribution to the nearest target point pairs according to Euclidean criterion. In fact, determining the farthest point for each target point in second phase of the algorithm NCAR, extract the maximum nearest point hit to the target point so that the nearer points of the maximum nearest point hit to target point would no longer be examined and would be definitely among the neighbors of target point. In fact, the position of the farthest point is the point distinguished for two different groups neighborhood for two target points (based on Euclidean criterion). In this way, the complexity of the algorithm decreases.

Suppose $M_i \in FP_t$ which applies to equation 10 and 11, and it is inserted in $k_{1t} = \frac{d(T_t, M_i)}{d(T_{t+1}, M_i)}$. Then, in order to separate neighborhood points around the target points, the following three states are obtained:

(1) $k_{1t} > 1$     (2) $k_{1t} < 1$     (3) $k_{1t} = 1$

(1) $k_{1t} > 1$: So by eliminating $FP_t$ Members from $M$ set, we obtain a new set of $FP_t$ are the members of which apply to equation 10 and 11. Then $k_{2t} = \frac{d(T_t, M_i)}{d(T_{t+1}, M_i)}$ is inserted. If we continue this trend, we reach the emphatically ascending sequence of $k_{1t} < k_{2t} < k_{3t} < \cdots < k_{jt}$ in which $j \leq n$, and $n$ indicates the total number of $M$ data.

**Definition:** The mean of $k_{lt}$ s around the target points and inside Apollonius circles is shown by $\bar{k}_{lt}$ and is defined as:

$$\bar{k}_{lt} = \frac{k_{lt} + k_{(l+1)t} + \ldots + k_{jt}}{j + 1 - l} \qquad l \in \{1, 2, \ldots, j\} \tag{14}$$

Obviously the mean has an emphatically ascending sequence and has a tendency toward infinite around the target points so that it would form nearest target point neighborhoods in the ascending sequence.

**Definition:** The $k_{lt}$ variances obtained through the above-mentioned stages are shown with $s_{lt}^2$ and are defined as:

$$s_{lt}^2 = \begin{cases} \frac{\sum_{i=1}^{j} k_{it}^2 - \bar{k}_{lt}^2}{j - 1} & l \in \{1, 2, \ldots, j-1\} \\ 0 & l = j \end{cases} \tag{15}$$

It is obvious that $k_{1t} < k_{2t} < k_{3t} < \cdots < k_{jt}$ is an emphatically ascending sequence and $s_{1t}^2 > s_{2t}^2 > s_{3t}^2 > \ldots > s_{jt}^2$ or the variance of the points are emphatically descending around the target points with a tendency towards zero around the target points; the nearer the variance value is to zero, the stronger the relationship between the target points and its neighborhood becomes. In this case, for the points having $k_{1t} > 1$, $T_{t+1}$ target points are optimal.

(2) $k_{1t} < 1$: By eliminating $FP_t$ members from set M, a new set of $FP_t$ is obtained the members of which apply in equation 10 and 11, and $k_{2t} = \frac{d(T_t, M_i)}{d(T_{t+1}, M_i)}$.

By continuing this trend, we obtain the emphatically descending sequence of $k_{1t} > k_{2t} > k_{3t} > \cdots > k_{jt}$ in which $j \leq n$, and $n$ refers to the total number of $M$ data. The mean of $k_{1t}$ s around the target points and inside the Apollonius circles are defined using Eq. (14). In this state, the mean has an emphatically descending sequence with a tendency towards zero around the target points.

Moreover, the $k_{1t}$ variances around the target points and inside the Apollonius circles are defined using Eq. (15). In this state, the variance has an emphatically descending sequence with a tendency towards zero around the target points.

In this state, for the points with $k_{1t} < 1$, $T_t$ target points are optimal.

(3) $k_{1t} = 1$: Therefore, $d(T_t, M_i) = d(T_{t+1}, M_i)$ indicating that all $FP_t$ members from set $M$ are located on the perpendicular bisector of the line segment target point of $T_t, T_{t+1}$. Now by eliminating $FP_t$ members of set $M$ and by considering equation 10 and 11, we have $k_{2t} = \dfrac{d(T_t, M_i)}{d(T_{t+1}, M_i)}$. It is clear that $k_{2t} \neq 1$ indicating that either state 1 or 2 will be repeated.

### 3.4. Computational complexity of the NCAR algorithm

If $n$ is the all number of points in the data sets, the worst complexity time of step 1 for NCAR algorithm occurs O (n2) in calculating the distance matrix. Step 2 checks the furthest point of target points and quicksort in the set related to any target point respectively cost O (n2) and O (nlogn). The empirical analysis displays that step 3 has O (n2) in calculating the similarity matrix. The worst complexity time of GG, RNG, NC and DPC-PCA are O (n3) which is the density calculation (İnkaya et al., 2015, Yang et al., 2016, Due et al., 2016, Wang et al., 2016).

## 4. Assessment Criteria

Different criteria have been proposed in different studies to compare accuracy of finding data point neighborhood among several algorithms. If neighborhood criterion is defined according to its precision in clustering and internal similarity among the neighbors, and the distance between the points is highly significant in the neighborhood set, the assessment criteria suggested by Rand (1971) including Variability Neighborhood (VN) (Eq. (16)) and Similarty Neighborhood (SN) (Eq. (17)) can be used.

$$VN = \frac{1}{D'} \sum_{i \in D'} \left( \frac{\sum_{j \in CS_i} (d_{ij} - d_i')^2}{|CS_i|} \right)^{\frac{1}{2}} \qquad (16)$$

In Eq. 16, $D' = \{i : |CS_i| \geq 2\}$.

$$SN = \frac{\sum_{i \in D} SN_i}{|D|} \qquad (17)$$

SN studies the similarity inside the neighbors, and if (SN $i = 1$) it means that the points in cluster $i$ belong to the same target cluster, and if (SN $i = 0$), the points $n$ cluster $i$ do not belong to the same target cluster. Rand index (RI) is used to penalize clusters splits and merges (Eq. (18)). The high value of RI displays that algorithm recover the original label class correctly. It refers to the proportion of the number of the points which the system has labeled correctly to the total number of the points the system has labeled.

$$RI = \frac{a+d}{a+b+c+d} \tag{18}$$

a= the number of pair points with the same cluster label and the same class label.

b= the number of pair points with the same cluster label and different class label.

c= the number of pair points with the diffrent cluster label and same class label.

d= the number of pair points with the diffrent cluster label and diffrent class label.

The algorithm NCAR with SOM, ε-neighborhood, NC, and KNN are compared. In KNN the value of *k* is set to 5% and 10% of the points in the data set, where we donate these configurations by KNN1 and KNN2, respectively. The purpose is to make sure that the value of *k* is least the size of the smallest cluster. Inkaya et al., (2015) procedure is used to set the value of ε and *k*. *k*-distance graph (*k*=4) is used to determine *ε* value for each data set. The procedure used for interpreting SOM by Inkaya et al., (2015) was applied. Three settings were tested in which a number of neurons are determined through one of the following methods: i. In order to test SOM clustering accuracy, the number of target structures is used. ii. NNCAR is used to determine the number of for aim to compare with NCAR. iii. MN ,the purpose of testing the capability of SOM for constructing clusters, the maximum number of used clusters (MN = $\sqrt{n}$) (Vesanto & Alhoniemi, 2000). These settings are SOM-NTC, SOM-NNCAR and SOM-MN, respectively.

## 5. Experimental study

We presented a comparative study of NCAR with other competing neighborhood construction and clustering algorithms (Inkaya et al., 2015, Due et al., 2016, Wang et al., 2016). NCAR algorithm gives emphasis on the local specification and the geometry specification in data sets. NCAR algorithm is like poles attracting similar data sets according to the structure geometry Apollonius circles. Two groups of data sets include real data sets and artificial data sets are used:

Group 1 data sets: The first group include nine real data sets (UCI Machine Learning Repository) which are used in our simulation experiments in order to validate the performance of NCAR algorithm in comparison with other algorithms (Blake & Merz, 1998). The detailed characteristics of these data sets are provided in Table 2. The algorithm NCAR is compared to state-of-the-art and well-known algorithms.

Group 2 data sets: The second group includes 45 artificial data sets which are used in our simulation experiments in order to validate the performance of NCAR algorithm in comparison with other algorithms (İnkaya et al., 2015) . These data sets were obtained from two sources (İnkaya et al., 2015, Liu et al 2008, Nosovskiy et al., 2008). The largest data sets has 2100 points, ten outliers and various shapes. The second group is composed of higher and two dimensional data sets. For example, data-c-cc-nu-n has circular and elongated shape. This data set has intercluster and intracluster density differences.

**Table 2**
Details of real-world data sets taken from UCI.

| Data set | #instance | #feature | #cluster |
|---|---|---|---|
| Iris | 150 | 4 | 3 |
| Wine | 178 | 13 | 3 |
| Heart | 270 | 13 | 2 |
| Waveform | 5000 | 21 | 3 |
| Sonar | 208 | 60 | 2 |
| Glass | 214 | 10 | 6 |
| Pen-based digits | 109,962 | 16 | 10 |
| LED digits | 500 | 7 | 3 |
| Seeds | 210 | 7 | 3 |

## 5.1 Experiments on real data sets

The resulting neighborhood constructed by the NCAR algorithm is reported on real data sets in Table 3. Our experiments were performed to evaluate the performance of the NCAR algorithm with several state-of-the-art algorithms including, DPC-KNN and DPC-PCA algorithms on nine real-world data sets. Similar results were obtained based on RI metric in Table 3. For example the accuracy of the NCAR algorithm for the wine data set was 0.90, while accuracy of DPC-KNN and DPC-PCA algorithms were 0.73 and 0.74, respectively. In Table 3, p is parameter which is used to define the number of neighbors in the first step process in NCAR, DPC-KNN and DPC-PCA algorithms. In our algorithm, we select the parameter p from fixed value [5%]. In DPC-KNN and DPC-PCA, the parameter p is also selected from [2% to 30%]. The experimental results demonstrated that the NCAR algorithm performed better than DPC-KNN and DPC-PCA in most data sets. The DPC-KNN are not sensitive to the local geometric and do not perform well when data points have relatively high dimensions. The DPC-PCA reduces the data and keeps 99% of principal components to eigenvectors.

The outputs of Table 3 show that DPC-KNN algorithm has better performance in the data sets having lower dimensions since it does not lose any information about the features. The DPC-PCA is an algorithm based on principal component analysis that reduces the data and keeps 99% of principal components to eigenvectors. However, DPC-PCA algorithm does not perform well on the data set with lower dimensions. On the contrary, it works better on the data sets with higher dimensions; the reason lies in the fact that it eliminates the information about less important features, and it shows higher identification accuracy on the remaining dimensions than the similar sets. The ability of the NCAR algorithm to identify neighborhood groups is better in lower dimensions than for high-dimensional data sets. For example, in Sonar, Waveform, and Pen-based digits data sets, the accuracy of our algorithm is less than the other algorithms. Fig. 9, simply show neighborhood construction of NCAR algorithm in data sets seeds of Group 1 with reassign points out of radius Apollonius circle and overlap.

**Table 3**
Comparison of RI values provided by different clustering methods.

| Data sets | Algorithms | | |
|---|---|---|---|
| | DPC-KNN | DPC-PCA | NCAR |
| Iris | 0.9123 | 0.9123 | **0.9317** |
| | P=9% | P=8% | **P=5%** |
| Wine | 0.7360 | 0.7403 | **0.9081** |
| | P=14% | P=0.6% | **P=5%** |
| Heart | 0.5642 | **0.5669** | 0.5556 |
| | P=5% | **P=5%** | P=5% |
| Waveform | 0.6720 | **0.7386** | 0.7333 |
| | P=0.2% | **P=0.1%** | P=5% |
| Sonar | 0.5221 | 0.5066 | 0.4943 |
| | P=4% | P=3% | P=5% |
| Glass | 0.8146 | 0.8653 | **0.8865** |
| | P=1% | P=30% | **P=5%** |
| Pen-based digits | 0.7618 | **0.7623** | 0.7034 |
| | P=1% | **P=0.2%** | P=5% |
| Seeds | 0.9143 | 0.9143 | **0.9254** |
| | P=2% | P=2% | **P=5%** |
| LED digits | 0.7460 | 0.6700 | **0.7691** |
| | P=6% | P=6% | **P=5%** |

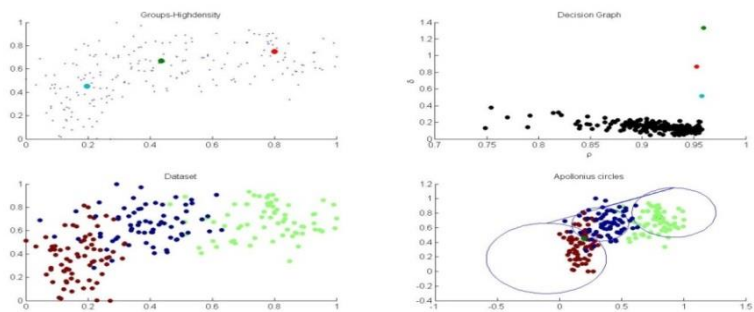

**Fig. 9.** Neighborhood for data set Seeds of Group 1 with the NCAR algorithm with Reassign points out of radius Apollonius circle and overlap.

## 5.2 Experiments on artificial data sets

The resulting neighborhood constructed by the NCAR algorithm and the other methods are reported on 45 artificial data sets in Table 4 and Table 5. Experimental results by SOM-NTC, SOM-NNCAR and SOM-MN are reported in Table 4. Among 45 data sets SOM-NTC and SOM-NNCAR find the target clusters in 3, 2, 8, and 21 data sets, but SOM-MN cannot extract the target clusters in any of the data sets, and NCAR found most of clusters correctly.

**Table 4**
Comparison results of SOM-NTC, SOM-NNC and SOM-MN in terms of mean accuracy,
Similarity and variability for Group 2 on 45 data sets (the best performer for each algorithm is bolded).

| Accuracy and similarity | Algorithms | | |
|---|---|---|---|
| | SOM-NTC | SOM-NNCAR | SOM-MN |
| RI | | | |
| Mean | **0.81** | 0.71 | 0.45 |
| Min | **0.50** | 0.20 | 0.12 |
| SN | | | |
| Mean | 0.64 | 0.87 | **0.93** |
| Min | 0 | 0.08 | **0.16** |
| VN | | | |
| Mean | 0.14 | 0.06 | **0.06** |
| Time(s) | | | |
| Mean | **0.69** | 0.81 | 1.00 |
| Max | **2.34** | 2.69 | 2.31 |

Our algorithm is compared in terms of RI (accuracy), SN (similarity), and VN (variability) with KNN1, KNN2, ε-neighborhood (well-known algorithms) and NC (state-of-the-art algorithm) in Table 5. In group 2 of data sets, the NCAR algorithm yields best RI, VN, and SN values. Moreover, our algorithm performs better than all the competing methods in terms of RI significantly. Among 16 data sets with outliers, NCAR finds all outliers in whole data set, but SOM-NNC find outliers in a single data set and ε-neighborhood in two data sets. KNN cannot find the outliers in any of data sets. NCAR algorithm is successful in extracting the number of clusters correctly. Obviously, the proposed algorithm obtained the best performance in most of the cases. Table 5 compares NCAR algorithm with NC algorithm both of which use geometric based method. NCAR has high accuracy with artificial data and is takes less time than NC algorithm. However, ε- neighborhood, KNN1, and KNN2 take less time than NCAR, respectively. The higher speed of this algorithm in comparison with other well-known algorithms is due to finding similarity points out of the radius f Apollonius circle and the points between the two Apollonius circles in the last phase of the algorithm. NCAR algorithm forms new similar groups based on the Apollonius geometry obtained from the points which are an accurate ratio of the distance between the two target points and isolation of the nearest neighborhood points between them. The data set R15 of Group 2 with 15 target points, 600 data points is used to illustrate the performance of the NCAR algorithm without reassigning points out of radius of Apollonius circles and inside overlap of Apollonius circles. R15 is generated as 15 similar 2-D Gaussian distributions that are positioned in rings. The analysis of R15 is presented in Fig. 10. Fig. 11, simply show neighborhood construction of NCAR algorithm in train2 of Group 2 (mean Runtime(s) = 0.23, RI = 0.9618).

**Table 5**
Comparison results of KNN1, KNN2, ε-neighborhood, NC and NCAR in terms of mean accuracy, Similarity and variability for Group 2 on 45 data sets (the best performer for each algorithm is bolded).

| Accuracy and similarity | Algorithms | | | | |
|---|---|---|---|---|---|
| | KNN1 | KNN2 | ε-neighborhood | NC | NCAR |
| RI | | | | | |
| Mean | 0.78 | 0.74 | 0.78 | 0.90 | **0.92** |
| Min | 0.30 | 0.30 | 0.30 | 0.69 | **0.71** |
| SN | | | | | |
| Mean | 0.96 | 0.90 | 0.88 | **0.99** | **0.99** |
| Min | 0.83 | 0.83 | 0 | 0.93 | **0.94** |
| VN | | | | | |
| Mean | 0.09 | 0.15 | 0.15 | 0.10 | **0.09** |
| Time(s) | | | | | |
| Mean | 6.31 | 7.95 | **5.37** | 12.84 | 7.77 |
| Max | 22.04 | 28.62 | **19.28** | 43.38 | 21.08 |

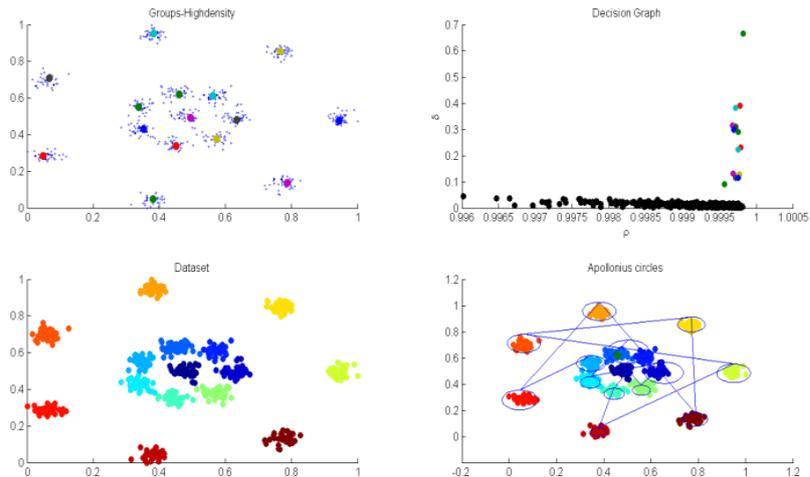

**Fig. 10.** Neighborhood for 15 target points and 600 data points in data set R15 (mean Runtime(s) = 0.39, RI = 0.9991) with the NCAR algorithm without Reassign points out of radius Apollonius circle and overlap.

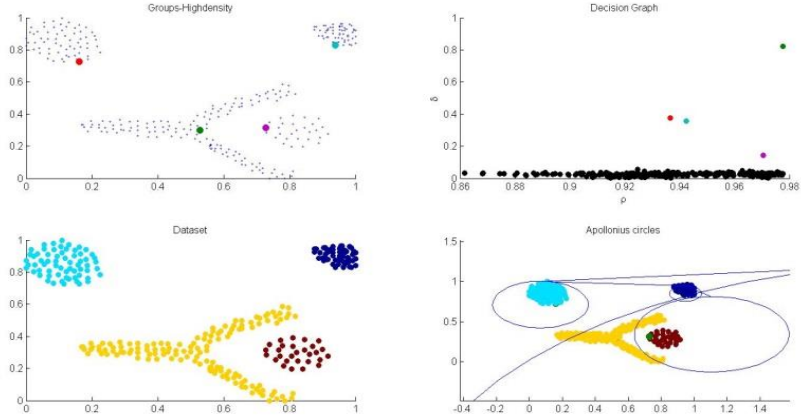

**Fig. 11.** Neighborhood for data set train2 of Group 2 (mean Runtime(s) = 0.23, RI = 0.9618) with the NCAR algorithm with Reassign points out of radius Apollonius circle and overlap.

We proposed a three-step clustering algorithm, namely NCAR, for data sets with spatial characteristics. The algorithm includes neighborhood construction, outlier detection, and construction of final clusters phases, which combine density and geometry information with graph theoretic concepts like proximity and connectivity. The criteria defined for outlier detection and construction of final clusters are based on the assessment of the Apollonius lengths relative to their neighborhoods in order to handle arbitrary shapes and density variations in distance calculations. We evaluated the performance of the algorithm using data sets having various properties, and we conducted a benchmark analysis with some well-known competing approaches. The numerical results indicated that the proposed algorithm is capable of finding clustering solutions close to the target clusters with arbitrary shapes and different densities when the intercluster distances are larger than the intracluster distances. NCAR has a favorable performance, and low compelxity time compared to the other geometry-based and density-based methods. In previous studies only proximity and direction information is taken into account to explore the local properties in constructing neighborhoods. However, these are not sufficient when there are density variations in a local region. As a remedy density-based connectivity through Apollonius is used in the proposed NCAR algorithm. The experimental results indicate that the idea is effective and relatively more purified neighborhoods are constructed. As the number of cluster mixes is small in the purified neighborhood, it enables effectivity in clustering. Thus, NCAR can be used as a preprocessing step for a neighborhood-based clustering method. If the succeeding clustering algorithm is specialized in the merging operations, target clusters can be found with more ease and accuracy.

## 6. Conclusion

Neighborhood construction problem is an important issue and is applied in many engineering methods. Most of the proposed neighborhood construction models have focused on distance-based method and machine learning algorithms. Most of these methods are either parametric, or they do not have safe and flexible neighborhood area fitting the changes in the size of data points. In other words, they do not feature adaptability. In this study, we have proposed a new neighborhood construction algorithm called NCAR which runs on data sets without any prior knowledge. A new algorithm for solving the neighborhood construction problem has been developed on geometric position, geometric relationship and the distance between data points. The algorithm has been analyzed on several well-known real and artificial data sets, and experimental results are very encouraging. NCAR algorithm is more accurate than other algorithms up to almost 7-12%. Neighborhood construction in the proposed algorithm aims at forming locally similar neighborhood points.

Possible futurere search directions are as follows.

- NCAR give emphasis on the local characteristics among Apollonius circles. The focuse can be connection global Apollonius circles and analysis data points inside Apollonius circles.

- Another future research direction can be exploration of new the angle between data points on Apollonius circles for find good connectivity among them.
- Hybridization of NCAR with metaheuristic or PCA are useful methods for increase accuracy in finding neighborhood.
- Target points extraction with other density-based methods at the right moment along with computations is still a field calling for further explorations.
- NCAR will be using for recommendation system design and medical decision support system to predict disease in future.

## References

Awad, M. (2010). Segmentation of satellite images using Self-Organizing Maps. In Self-Organizing Maps: InTech.

Bautista, J. A. R., Cárdenas, S. L. C., Zavala, A. H., & Huerta-Ruelas, J. A. (2018). Review on plantar data analysis for disease diagnosis. *Biocybernetics and Biomedical Engineering*.

Blake, C. L., & Merz, C. J. (1998). UCI Repository of machine learning databases [http://www. ics. uci. edu/~ mlearn/MLRepository. html]. Irvine, CA: University of California. Department of Information and Computer Science, 55.

Cardinal, J., Collette, S., & Langerman, S. (2009). Empty region graphs. *Computational geometry, 42*, 183-195.

Crnovrsanin, T., Muelder, C. W., Faris, R., Felmlee, D., & Ma, K.-L. (2014). Visualization techniques for categorical analysis of social networks with multiple edge sets. *Social Networks, 37*, 56-64.

Du, M., Ding, S., & Jia, H. (2016). Study on density peaks clustering based on k-nearest neighbors and principal component analysis. *Knowledge-Based Systems, 99*, 135-145.

Güney, S., & Atasoy, A. (2012). Multiclass classification of n-butanol concentrations with k-nearest neighbor algorithm and support vector machine in an electronic nose. *Sensors and Actuators B: Chemical, 166*, 721-725.

García-Pedrajas, N., del Castillo, J. A. R., & Cerruela-García, G. (2017). A Proposal for Local k Values for k-Nearest Neighbor Rule. *IEEE transactions on neural networks and learning systems, 28*, 470-475.

Hajjar, C., & Hamdan, H. (2013). Interval data clustering using self-organizing maps based on adaptive Mahalanobis distances. *Neural Networks, 46*, 124-132.

Hoshen, J. (1996). The GPS equations and the problem of Apollonius. *IEEE Transactions on Aerospace and Electronic Systems, 32*, 1116-1124.

İnkaya, T. (2015a). A parameter-free similarity graph for spectral clustering. *Expert Systems with Applications, 42*, 9489-9498.

Güngör, E., & Özmen, A. (2017). Distance and density based clustering algorithm using Gaussian kernel. *Expert Systems with Applications, 69*, 10-20.

İnkaya, T. (2015b). A density and connectivity based decision rule for pattern classification. *Expert Systems with Applications, 42*, 906-912.

İnkaya, T., Kayalıgil, S., & Özdemirel, N. E. (2015). An adaptive neighborhood construction algorithm based on density and connectivity. *Pattern Recognition Letters, 52*, 17-24.

Liu, D., Nosovskiy, G. V., & Sourina, O. (2008). Effective clustering and boundary detection algorithm based on Delaunay triangulation. *Pattern Recognition Letters, 29*, 1261-1273.

Maillo, J., Ramírez, S., Triguero, I., & Herrera, F. (2017). kNN-IS: An Iterative Spark-based design of the k-Nearest Neighbors classifier for big data. *Knowledge-Based Systems, 117*, 3-15.

Mohammadi, M., Raahemi, B., Mehraban, S. A., Bigdeli, E., & Akbari, A. (2015). An enhanced noise resilient K-associated graph classifier. *Expert Systems with Applications, 42*, 8283-8293.

Nettleton, D. F. (2013). Data mining of social networks represented as graphs. *Computer Science Review, 7*, 1-34.

Nosovskiy, G. V., Liu, D., & Sourina, O. (2008). Automatic clustering and boundary detection algorithm based on adaptive influence function. *Pattern Recognition, 41*, 2757-2776.

Pan, X., Luo, Y., & Xu, Y. (2015). K-nearest neighbor based structural twin support vector machine. *Knowledge-Based Systems, 88*, 34-44.

Papa, J. P., Fernandes, S. E. N., & Falcão, A. X. (2017). Optimum-path forest based on k-connectivity: Theory and applications. *Pattern Recognition Letters, 87*, 117-126.

Park, H., & Basole, R. C. (2016). Bicentric diagrams: Design and applications of a graph-based relational set visualization technique. *Decision Support Systems, 84*, 64-77.

Partensky, M. B. (2008). The circle of Apollonius and its applications in introductory physics. *The Physics Teacher, 46*, 104-108.

Pedrycz, W. (2010). The design of cognitive maps: A study in synergy of granular computing and evolutionary optimization. *Expert Systems with Applications, 37*, 7288-7294.

Pourbahrami, S. (2017). Optimization of LEACH method using distribute nodes and cell clustering. In Knowledge-Based Engineering and Innovation (KBEI), 2017 IEEE 4th International Conference on (pp. 0083-0087): IEEE.

Qin, Y., Yu, Z. L., Wang, C.-D., Gu, Z., & Li, Y. (2018). A Novel clustering method based on hybrid K-nearest-neighbor graph. Pattern Recognition, 74, 1-14.

Rand, W. M. (1971). Objective criteria for the evaluation of clustering methods. *Journal of the American Statistical association, 66*, 846-850.

Rodriguez, A., & Laio, A. (2014). Clustering by fast search and find of density peaks. *Science, 344*, 1492-1496.

Li, Z., & Tang, Y. (2018). Comparative density peaks clustering. *Expert Systems with Applications, 95*, 236-247.

Stork, D. G., Duda, R. O., Hart, P. E., & Stork, D. (2001). Pattern classification. *A Wiley-Interscience Publication*.

Varma, M. K. S., Rao, N., Raju, K., & Varma, G. (2016). Pixel-based classification using support vector machine classifier. In *Advanced Computing (IACC), 2016 IEEE 6th International Conference on* (pp. 51-55): IEEE.

Vehlow, C., Beck, F., & Weiskopf, D. (2015). The state of the art in visualizing group structures in graphs. In *Eurographics Conference on Visualization (EuroVis)-STARs* (Vol. 2): The Eurographics Association.

Veltkamp, R. C. (1992). The γ-neighborhood graph. *Computational geometry, 1*, 227-246.

Vesanto, J., & Alhoniemi, E. (2000). Clustering of the self-organizing map. *IEEE Transactions on neural networks, 11*, 586-600.

Wang, M., Zuo, W., & Wang, Y. (2016). An improved density peaks-based clustering method for social circle discovery in social networks. *Neurocomputing, 179*, 219-227.

Yang, Y., Han, D.-Q., & Dezert, J. (2016). An angle-based neighborhood graph classifier with evidential reasoning. *Pattern Recognition Letters, 71*, 78-85.